\documentclass[journal=apchd5,manuscript=article]{achemso}

\usepackage[version=3]{mhchem} % Formula subscripts using \ce{}
\usepackage{amsmath}

\author{Giovanni Pellegrini}
\affiliation{Dipartimento di Fisica, Politecnico di Milano, Piazza Leonardo da Vinci 32, I-20133 Milano, Italy}
\email{giovanni.pellegrini@polimi.it}
\author{Marco Finazzi}
\affiliation{Dipartimento di Fisica, Politecnico di Milano, Piazza Leonardo da Vinci 32, I-20133 Milano, Italy}
\author{Michele Celebrano}
\affiliation{Dipartimento di Fisica, Politecnico di Milano, Piazza Leonardo da Vinci 32, I-20133 Milano, Italy}
\author{Lamberto Du\`{o}}
\affiliation{Dipartimento di Fisica, Politecnico di Milano, Piazza Leonardo da Vinci 32, I-20133 Milano, Italy}
\author{Maria Antonia Iat\`{i}}
\affiliation{CNR-IPCF, Istituto per i Processi Chimico-Fisici, viale F. Stagno D'Alcontres 37, I-98166 Messina, Italy}
\author{Onofrio M. Marag\`{o}}
\affiliation{CNR-IPCF, Istituto per i Processi Chimico-Fisici, viale F. Stagno D'Alcontres 37, I-98166 Messina, Italy}
\author{Paolo Biagioni}
\affiliation{Dipartimento di Fisica, Politecnico di Milano, Piazza Leonardo da Vinci 32, I-20133 Milano, Italy}
\email{paolo.biagioni@polimi.it}

\title[]{Superchiral Surface Waves for All-Optical Enantiomer Separation}


\begin{document}

% \begin{tocentry}

% Some journals require a graphical entry for the Table of Contents.
% This should be laid out ``print ready'' so that the sizing of the

% Inside the \texttt{tocentry} environment, the font used is Helvetica
% 8\,pt, as required by \emph{Journal of the American Chemical

% The surrounding frame is 9\,cm by 3.5\,cm, which is the maximum
% permitted for  \emph{Journal of the American Chemical Society}
% graphical table of content entries. The box will not resize if the
% content is too big: instead it will overflow the edge of the box.

% This box and the associated title will always be printed on a
% separate page at the end of the document.

% \end{tocentry}

\begin{abstract} We introduce the use of superchiral surface waves for the all-optical separation of chiral compounds. Using a combination of
electrodynamics modeling and analytical techniques, we show that the
proposed approach provides chiral optical forces two orders of magnitude larger
than those obtained with circularly polarized plane waves. Superchiral surface
waves allow for enantiomer separation on spatial, temporal and size scales
than would not be achievable with alternative techniques, thus representing a
viable route towards all-optical enantiomer separation. \end{abstract}

Chirality is the geometrical property of three-dimensional bodies that are
distinct from their mirror image. Enantiomers, i.e. objects that display
opposite chirality, have in common most of their physical and chemical
properties, and must interact with a chiral environment to exhibit their chiral
attributes. Since chirality is pervasive in nature, and likewise in a large
variety of biomolecules, the vast majority of biochemical processes is strongly
influenced by the chiral properties of the involved chemical compounds
\cite{Meierhenrich_2008, Fasman_1996}. For these reasons the analysis and
separation of chiral molecules has gained traction in the biochemical and
pharmaceutical industries. Currently, pharmaceutical manufacturing is moving
towards the production of enantiopure chiral molecules, with the expectation
that the vast majority of drugs would be chiral by 2020
\cite{noauthor_development_1992,noauthor_chiral_2012}. Circular dichroism (CD)
spectroscopy, i.e. the measurement of differential absorption of left and right
circularly polarized light, is one of the reference techniques when dealing with
enantiomer discrimination. Nevertheless, the differential nature of CD signals
tipically leads to low signal to noise ratios, thus complicating the measurement
of small amount of chiral molecules. In this scenario, new optical
characterization techniques have been proposed, which are based on the
introduction of superchiral electromagnetic fields, i.e. fields that display an
optical chirality $C=\frac{\varepsilon_{0}\omega}{2}\mathrm{Im}\{\mathbf{E^{*}
\cdot B}\}$ larger than that of a circularly polarized plane wave
\cite{barron_molecular_2004,Tang_2010,Tang_2011}. A variety of solutions, based
on plasmonic and photonic nanostructures, have been designed to obtain
superchiral light fields exploitable in realistic scenarios, with a particular
focus on designing solutions capable of analyzing small amounts of chiral
molecules
\cite{Sch_ferling_2012,Sch_ferling_2014,Nesterov_2016,Abdulrahman_2012,Sch_ferling_2016,Govorov_2010,Govorov_2012,Hendry_2012,Sch_ferling_2012a,Valev_2013,Lu_2013,Frank_2013,Liu_2014,Valev_2014,Tullius_2015,Finazzi_2015}.
In parallel with the search for efficient superchiral sensing platforms, the
research community devoted significant attention to the study and design of
electromagnetic fields for the generation of enantioselective optical forces
\cite{canaguier-durand_mechanical_2013,cameron_discriminatory_2014,cameron_diffraction_2014,tkachenko_helicity-dependent_2014,tkachenko_optofluidic_2014,wang_lateral_2014,alizadeh_transverse_2015,alizadeh_dominant_2016,hayat_lateral_2015,rukhlenko_completely_2016,zhao_enantioselective_2016,zhao_nanoscopic_2017,zhang_all-optical_2017,cipparrone_gabriella_chiral_2011,donato_light-induced_2016,donato_polarization-dependent_2014,brzobohaty_chiral_2016,nieto-vesperinas_optical_2010},
i.e. optical forces that can trap, separate, or more in general discriminate
between the two different enantiomers of a  chiral chemical compound. In this
context, several promising approaches have been proposed to tackle the problem.
As an example, both evanescent waves (EW) and incoherent combinations of
circularly polarized plane waves (PW) are capable of generating purely chiral
enantioselective optical forces, whose magnitude and direction depend on
the chiral polarizability $\chi$ of the target chiral compound
\cite{hayat_lateral_2015,rukhlenko_completely_2016}. On a different note,
plasmonic tweezers can serve as chiral optical traps, where the presence of
enantioselective forces has been experimentally verified
\cite{zhao_enantioselective_2016,zhao_nanoscopic_2017}. Nevertheless, despite
the numerous efforts, up to now the magnitude of the generated optical forces
only allows to manipulate chiral particles that are significantly larger than
typical proteins and pharmaceutically relevant molecules.

Here we show that, by exploiting a 1-dimensional photonic crystal (1DPC) capable
of sustaining Superchiral Surface Waves (SSWs) \cite{pellegrini_chiral_2017}, we
can separate chiral particles with radii $r$ and chiral polarizability $\chi$ as
small as $r \sim 5$~nm and  $\frac{c \, |\chi|}{4 \pi} \sim 10^{-21}$~cm$^{3}$
on spatial scales of tens of micrometers and temporal scales of tens of seconds,
reaching enantionpurity above 99$\%$ with the advantage of a platform that is
inherently compatible with microfluidics setups and applications.\

When working in dipolar approximation, we can express the force exerted by a
monochromatic electromagnetic field on a chiral dipole as $\mathbf{F} =
\mathbf{F_{0}} + \mathbf{F_{\mathrm{int}}}$ where, in S.I. units, we have
\cite{hayat_lateral_2015}
\begin{equation}
\begin{split}
\mathbf{F_{0}} & =  \frac{\mathrm{Re}\{\alpha_{\mathrm{e}}\}}{\varepsilon_{0}\varepsilon}\nabla u_{\mathrm{e}} + \frac{\mathrm{Re}\{\alpha_{\mathrm{m}}\}}{\mu_{0}\mu}\nabla u_{\mathrm{m}} - c \, \omega \mathrm{Re}\{\chi\}\nabla h \\
& + 2 \omega \Big( \frac{\mu}{\varepsilon_{0}} \mathrm{Im}\{\alpha_{\mathrm{e}}\} \mathbf{p_{e}^{o}} + \frac{\varepsilon}{\mu_{0}} \mathrm{Im}\{\alpha_{\mathrm{m}}\} \mathbf{p_{m}^{o}} \Big) - c^{2} \, \mathrm{Im}\{\chi\}[\nabla \times \mathbf{p}-2 k^{2}\mathbf{s}]
\end{split}
\label{eq:chiral_force}
\end{equation}
and
\begin{equation}
\begin{split}
\mathbf{F_{int}}&= -\frac{c^{3}k^{4}}{6 \pi}
\Big(\mathrm{Re}\{\alpha_{\mathrm{e}}\alpha_{\mathrm{m}}^{*}\}\mathbf{p} -
				 \mathrm{Im}\{\alpha_{\mathrm{e}}\alpha_{\mathrm{m}}^{*}\}\mathbf{p''} + |\chi|^{2}\mathbf{p}\Big) \\
&-\frac{c^{2}k^{5}}{3 \pi n}
\Big(\frac{\varepsilon}{\mu_{0}} \mathrm{Re}\{\chi\alpha_{\mathrm{m}}^{*}\}\mathbf{s_{m}}
+ \frac{\mu}{\varepsilon_{0}} \mathrm{Re}\{\chi\alpha_{\mathrm{e}}^{*}\}\mathbf{s_{e}}\Big).
\end{split}
\label{eq:int_force}
\end{equation}
Here $\omega$ is the light angular frequency, $c$ is the speed of light in
vacuum, $\varepsilon_{0}$ and $\mu_{0}$ are the vacuum permittivity and
permeability, $\varepsilon$ and $\mu=1$ are the relative permittivity and
permeability, $\alpha_{\mathrm{e}}$, $\alpha_{\mathrm{m}}$ and $\chi$ are the
electric, magnetic and chiral polarizability, $n=\sqrt{\varepsilon}$ is the
refractive index and $k=n \omega/c$ is the light wavevector. The equation terms
$u_{\mathrm{e}}=\frac{\varepsilon_{0}\varepsilon}{4}|\mathbf{E}|^{2}$,
$u_{\mathrm{m}}=\frac{\mu_{0}\mu}{4}|\mathbf{H}|^{2}$ and $h=\frac{1}{2\omega
c}\mathrm{Im}\{\mathbf{E} \cdot \mathbf{H}^{*} \}$ are the electric and magnetic
contribution to the energy density and the field helicity. The momentum density $\mathbf{p}$, its imaginary adjoint $\mathbf{p}''$ and its orbital electric and magnetic components $\mathbf{p_{e}^{o}}$ and $\mathbf{p_{e}^{o}}$ are defined as:
\begin{equation}
\begin{split}
\mathbf{p} & = \frac{1}{2c}\mathrm{Re}\{ \mathbf{E} \times \mathbf{H}^{*} \} \\
\mathbf{p''} & = \frac{1}{2c}\mathrm{Im}\{ \mathbf{E} \times \mathbf{H}^{*} \} \\
\mathbf{p_{e}^{o}} & = \frac{\varepsilon_{0}}{4 \omega \mu}\mathrm{Im}\{ \mathbf{E} (\mathbf{E} \otimes  \mathbf{E}^{*}) \} \\
\mathbf{p_{m}^{o}} & = \frac{\mu_{0}}{4 \omega \varepsilon}\mathrm{Im}\{ \mathbf{H} (\mathbf{H} \otimes  \mathbf{H}^{*}) \}.
\end{split}
\label{eq:orbital}
\end{equation}
where the $\otimes$ symbol stands for the tensor product. Finally we have the spin angular momentum $\mathbf{s}$ where:
\begin{equation}
\mathbf{s} = \mathbf{s_{e}} + \mathbf{s_{m}}  = -\frac{\varepsilon_{0}}{4 i \mu \omega} \mathbf{E} \times \mathbf{E}^{*} - \frac{\mu_{0}}{4 i \varepsilon \omega} \mathbf{H} \times \mathbf{H}^{*}.
\label{eq:spin}
\end{equation}
In this framework $\mathbf{F_{0}}$ represents the contribution coming from the
interaction between the electromagnetic field and the induced dipole moments,
while the $\mathbf{F_{int}}$ term arises from the interaction between the
induced electric and magnetic dipole moments. When looking at the expression of
the optical forces, it is clear that only the terms dependent on the chiral
polarizability $\chi$ can be used to obtain enantioselective forces. The adopted
mechanism rests on the simple idea that enantiomers of different handedness
display chiral polarizabilities of opposite sign, and thus will be subject to
optical forces with opposite directions. This leaves us with a limited number of
exploitable force terms, and in particular the gradient term $\mathbf{F_{g}}=c \,
\omega \mathrm{Re}\{\chi\}\nabla h$, the radiation pressure term
$\mathbf{F_{p}}=c^{2} \, \mathrm{Im}\{\chi\}[\nabla \times \mathbf{p}-2
k^{2}\mathbf{s}]$ and the chiral electric interaction term
$\mathbf{F^{e}_{int}}=\frac{c^{2}k^{5}}{3 \pi n}\frac{\mu}{\varepsilon_{0}}
\mathrm{Re}\{\chi\alpha_{\mathrm{e}}^{*}\}\mathbf{s_{e}}$, where the
$\frac{c^{2}k^{5}}{3 \pi n}\frac{\varepsilon}{\mu_{0}}
\mathrm{Re}\{\chi\alpha_{\mathrm{m}}^{*}\}\mathbf{s_{m}}$ contribution has been
discarded  because in usual conditions $|\alpha_{\mathrm{m}}|/\mu_{0} \ll
|\alpha_{\mathrm{e}}|/\varepsilon_{0}$ \cite{rukhlenko_completely_2016}.
Interestingly, each of these force terms has been employed in the literature to
devise efficient schemes where pure enantioselective forces are dominant: the
gradient terms in the case of plasmonic optical tweezers
\cite{zhao_enantioselective_2016}, the radiation pressure terms with the
incoherent sum of circularly polarized plane waves
\cite{rukhlenko_completely_2016} and the electric interaction term with
evanescent waves excited at a prism air interface \cite{hayat_lateral_2015}.
While extremely promising, each of these approaches comes with the respective
drawbacks. The working conditions of the plasmonic tweezers require that $c
\mathrm{Re}\{\chi\} \geq \mathrm{Re}\{\alpha\}_{\mathrm{e}}/\varepsilon_{0}$,
demanding the introduction of an index matching approach for the chiral particle
and the surrounding medium in order to minimize the magnitude of the electric
polarizability \cite{zhao_enantioselective_2016}. The electric interaction term
approach (EW scheme), while attractive for its simplicity, is better suited for
larger chiral particles, given the functional dependence on the electric
polarizability $\alpha_{\mathrm{e}}$ \cite{hayat_lateral_2015}. Finally, the
incoherent plane wave approach (PW scheme) allows to operate on large areas, yet
the lack of field enhancement mechanisms limits the magnitude of the obtainable
chiral forces for reasonable incident powers.
\begin{figure}[t!]
\begin{center}
\includegraphics[]{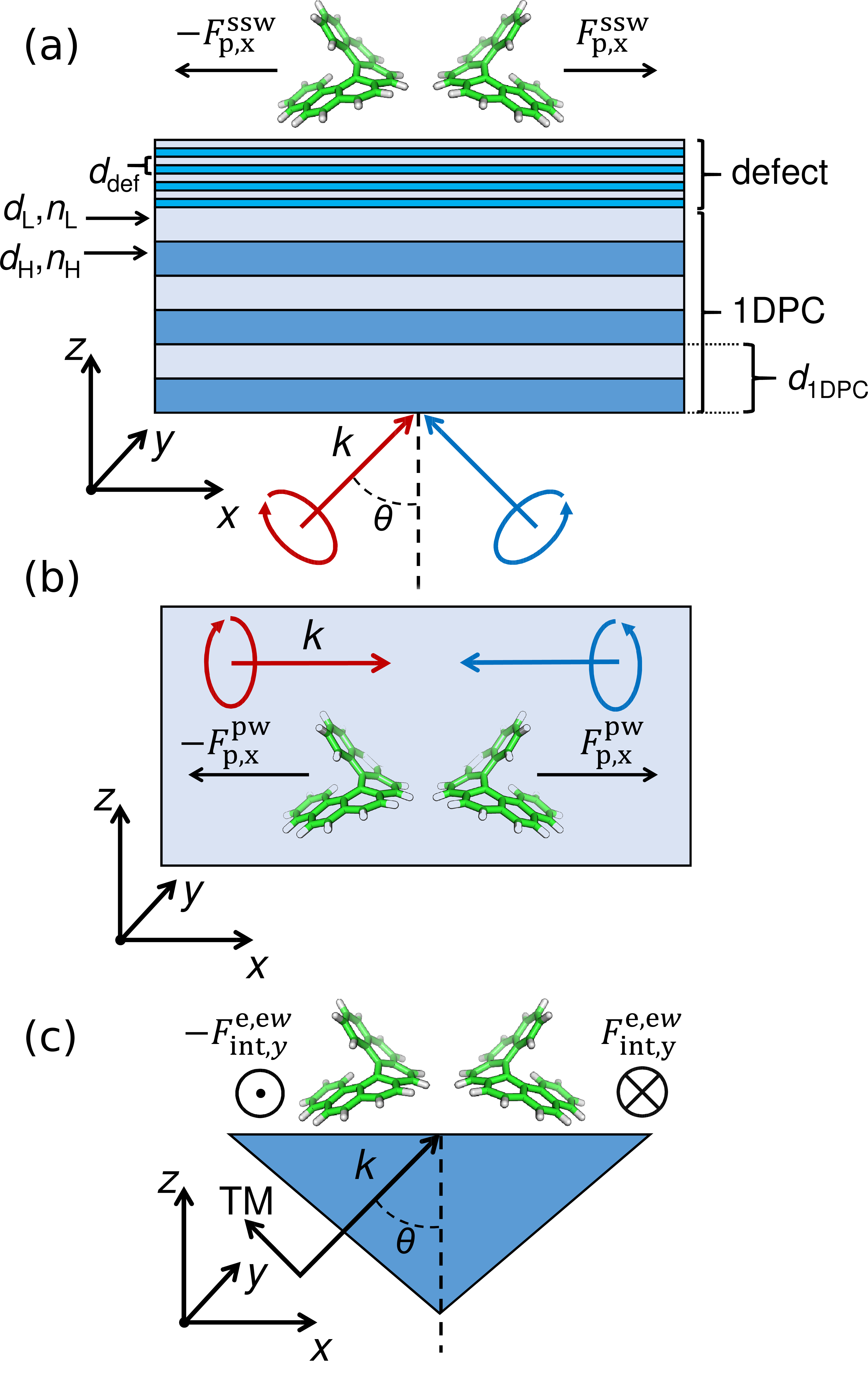}
\caption{\label{fig:scheme} A schematic representation of different setups for
the generation of optical enantionselective forces. (a) 1DPC SSW platform, with
the incoming incoherent elliptically polarized waves represented at the bottom,
and the corresponding enantiomeric separation happening at the top. The 1DPC and
termination periods are defined as $d_{\mathrm{1DPC}}$ and $d_{\mathrm{def}}$,
respectively, while $d_{\mathrm{H,L}}$ stands for the thickness of the high and
low refractive index materials in the 1DPC. (b) Incoherent counter-propagating
plane wave setup (PW). (c)  Evanescent wave setup (EW).}
\end{center}
\end{figure}
It would be ideal to devise a strategy that exploits the features of the PW
scheme, and at the same time the field enhancement properties typical of
plasmonic and photonic nanostructures. This would allow to obtain large and
uniform chiroptical forces over large surface areas. Unfortunately, localized
plasmonic and photonic resonances only provide field and chirality enhancements
over nanometric hot-spots and work in relatively narrow energy ranges, while
traditional surface waves such as Surface Plasmon Polaritons (SPPs) and Bloch
Surface Waves (BSWs) provide sizable field enhancements over large areas, but
cannot sustain the circular polarization state that is necessary to exploit the
chiral radiation pressure term $\mathbf{F_{p}}=c^{2} \,
\mathrm{Im}\{\chi\}[\nabla \times \mathbf{p}-2 k^{2}\mathbf{s}]$
\cite{maier_2007,Descrovi_2015}.

Here we propose a solution that simultaneously exploits the field enhancement
properties of photonic nanostructures and the large active areas typical of the
PW solution. We employ a radically different paradigm, which uses the optical
properties of 1DPCs to generate superchiral electromagnetic fields. The whole
approach rests on the idea that 1DPCs support both transverse-electric (TE) and
transverse-magnetic (TM) Bloch surface waves (BSWs) \cite{Sinibaldi_2014}. An
appropriate engineering of the multilayer structure, specifically the
introduction of an additional 1DPC termination with a much shorter lattice
parameter, allows for the superposition of the TE and TM dispersion relations.
\begin{figure}[t!]
\begin{center}
\includegraphics[]{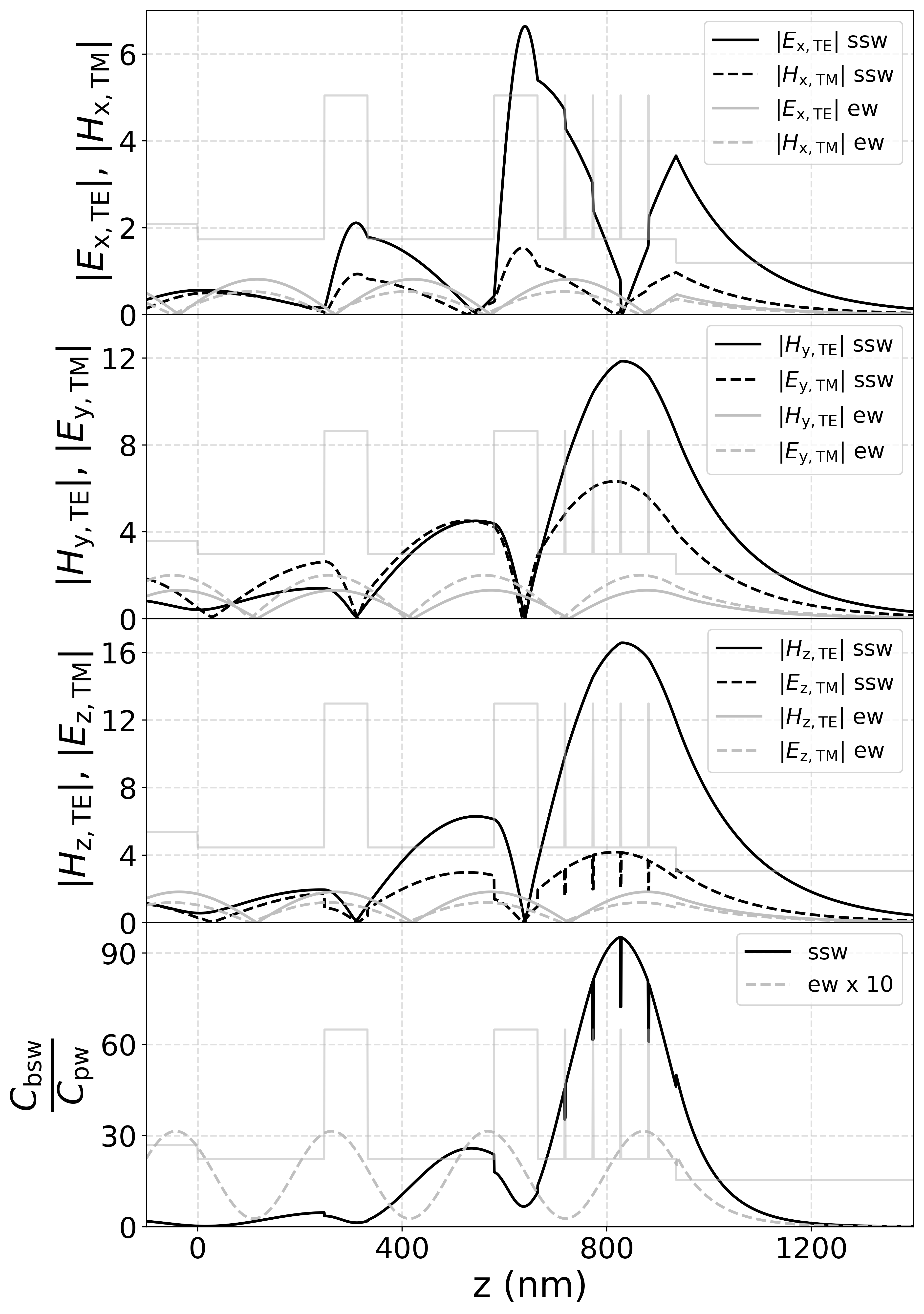}
\caption{\label{fig:fields} Local field and optical chirality plots along the
$z$-axis, across the 1DPC platform, for TE and TM incident plane waves, with a
$\lambda_{\mathrm{c}}=380$~nm wavelength and a
$\theta_{\mathrm{c}}\sim66^{\circ}$ incident angle. Plane waves are incident
from the left. (a) $E_{\mathrm{x}}$ and $H_{\mathrm{x}}$ components. (b)
$H_{\mathrm{y}}$ and $E_{\mathrm{y}}$ components. (c) $H_{\mathrm{z}}$ and
$E_{\mathrm{z}}$ components. (d) Optical chirality enhancement for an
elliptically polarized incident plane wave. The light gray field profiles are
calculated for an evanescent wave traveling at a prism-water interface
($n_{\mathrm{inc}}=1.53$, $n_{\mathrm{water}}=1.33$), where the prism and 1DPC
top interfaces have been aligned for a straightforward profile comparison.}
\end{center}
\end{figure}
This solution admits the simultaneous excitation of TE and TM surface waves, and
therefore the excitation of superchiral surface waves at the 1DPC surface, much
like the combination of two orthogonal linear polarization states with the
appropriate $\pi/2$ phase shift would result in a circular polarization state in
free space (Figure \ref{fig:fields}). This configuration leads to superchiral
surface waves that provide homogeneous, enhanced superchiral fields over
arbitrarily large surfaces and a wide spectral range
\cite{pellegrini_chiral_2017}, with optical chirality enhancements as large as
two orders of magnitude if compared with circularly polarized plane waves, as
already discussed in a previous publication. The same mechanism provides a
comparable intensification of the enantioselective forces, thus paving the road
towards all-optical enantiomer separation. For the following, we model the 1DPC
already described in Ref.\citenum{pellegrini_chiral_2017}. The 1DPC consists of
alternating  high (H) and low (L) refractive index materials, and in particular
we choose $\mathrm{Ta}_{2}\mathrm{O}_{5}$ ($n_{\mathrm{H}}=2.06+0.001i$) and
$\mathrm{SiO}_{2}$ ($n_{\mathrm{L}}=1.454+0.0001i$), while the upper
semi-infinite space is water ($n_{\mathrm{water}}=1.33$) and the incident medium
is a BK7 glass ($n_{\mathrm{inc}}$=1.53). The crystal periodicity is
$d_{\mathrm{1DPC}}=333$~nm, and the respective layer thicknesses are
$d_{\mathrm{H}}=f_{\mathrm{H}} \, d_{\mathrm{1DPC}}$ and
$d_{\mathrm{L}}=(1-f_{\mathrm{H}}) \, d_{\mathrm{1DPC}}$, where
$f_{\mathrm{H}}=0.26$ is the filling factor. The 1DPC termination is an
additional multilayer characterized by a period $d_{\mathrm{def}}$ much smaller
than $d_{\mathrm{1DPC}}$. It consists of $N_{\mathrm{def}}=5$ periods of
alternating $\mathrm{Ta}_{2}\mathrm{O}_{5}$ and $\mathrm{SiO}_{2}$ layers. The
total thickness is expressed as a function of the 1DPC parameters as
$t_{\mathrm{def}}=d_{\mathrm{L}} \, c_{\mathrm{def}}$, with
$c_{\mathrm{def}}=1.1$. The termination period then becomes
$d_{\mathrm{def}}=t_{\mathrm{def}}/N_{\mathrm{def}}$, and accordingly the layer
thicknesses are $d_{\mathrm{H,def}}=f_{\mathrm{H,def}} \, d_{\mathrm{def}}$ and
$d_{\mathrm{L,def}}=(1-f_{\mathrm{H,def}}) \, d_{\mathrm{def}}$, where
$f_{\mathrm{H,def}}=0.03$ is the termination filling factor. We finally note
that the termination ends with a low refractive index layer. We model the target
chiral compound as a spherical nanoparticle with optical constant
$\epsilon_{\mathrm{p}}=2.0(1+i)$, chirality parameter $\kappa=0.2$ and radii
varying from $r=2$~nm to $r=30$~nm
\cite{hayat_lateral_2015,zhao_enantioselective_2016}.
\begin{figure}[t!]
\begin{center}
\includegraphics[]{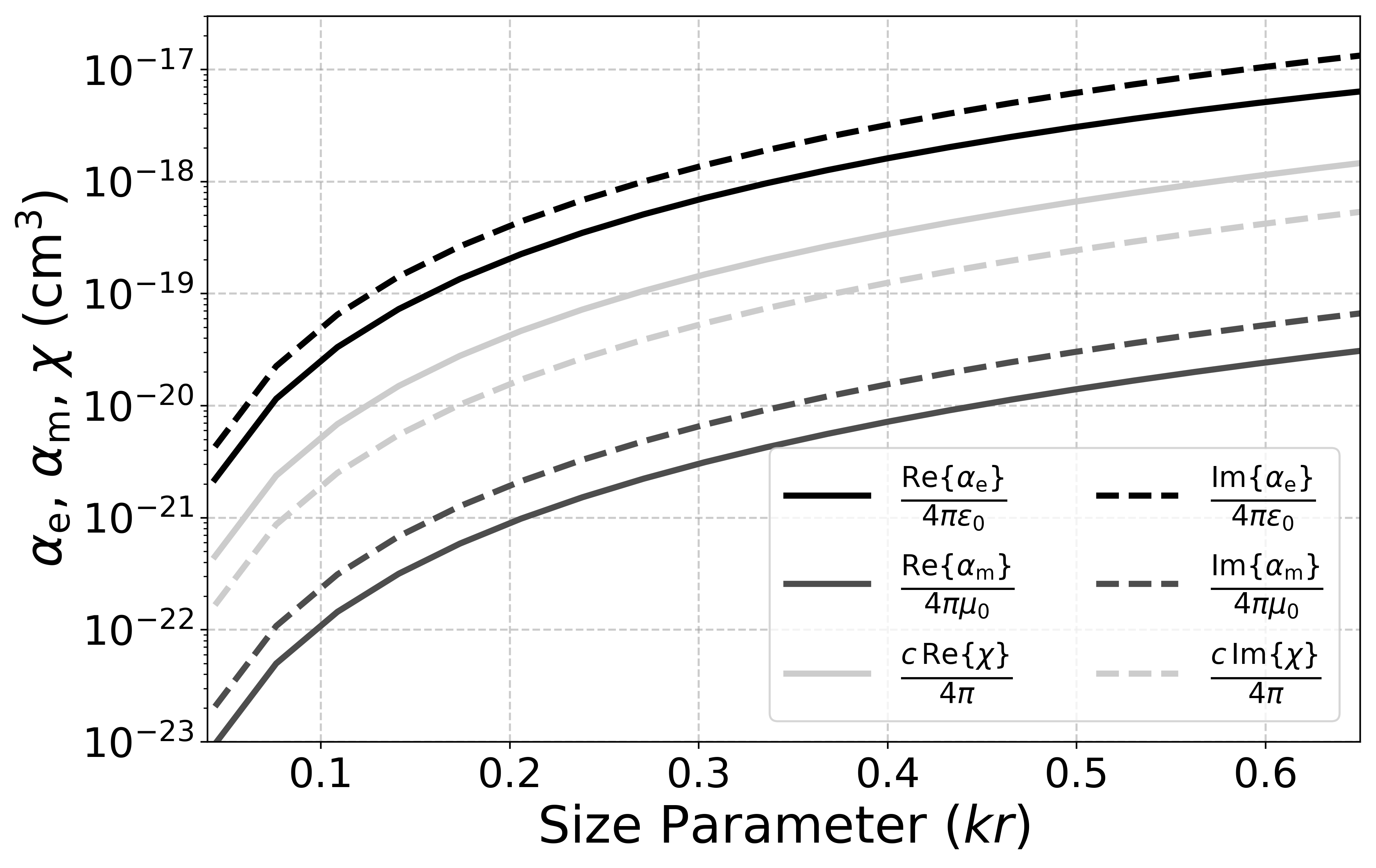}
\caption{\label{fig:pol} Electric, magnetic and chiral polarizability plot for a
chiral sphere with optical constant $\epsilon_{\mathrm{p}}=2.0(1+i)$, chirality
parameter $\kappa=0.2$ and radii varying from $r=2$~nm to $r=30$~nm. }
\end{center}
\end{figure}
Figure \ref{fig:pol} reports the chiral target polarizabilities as a function of
the size paramer $kr$ where $k=2\pi n_{\mathrm{water}}/ \lambda_{\mathrm{c}}$,
$\lambda_{\mathrm{c}}=380$~nm is the illumination wavelength and
$n_{\mathrm{water}}=1.33$ is the environment refractive index. It is clear how
the choice of these parameters preserves the typical magnitude hierarchy between
electric, chiral and magnetic polarizability, where we have
$|\alpha_{\mathrm{e}}|/\epsilon_{0} \gg c|\chi| \gg
|\alpha_{\mathrm{m}}|/\mu_{0}$ \cite{rukhlenko_completely_2016}. The chiral
polarizability values range from $\frac{c|\chi|}{4 \pi} \sim 10^{-21}$~cm$^{3}$
for $r=2$~nm to $\frac{c|\chi|}{4 \pi}\sim10^{-18}$~cm$^{3}$ for $r=30$~nm
particles, roughly keeping an order of magnitude gap with the electric and
magnetic polarizability terms within the whole size range.\

To analyze and put into context the chiroptical forces associated with SSWs, it
is useful to study them as a function of the chiral particle size, and compare
them to those obtainable with the already employed PW and EW schemes
\cite{hayat_lateral_2015,rukhlenko_completely_2016}. In the case of SSWs, we
employ the already described 1DPC structure illuminated at the coupling
wavelength and angle $\lambda_{\mathrm{c}}=380$~nm and $\theta_{\mathrm{c}} \sim
66^{\circ}$, and position the target $30$~nm above the 1DPC surface. For
simplicity, in the following we model a structure illuminated with a single
incident beam, keeping in mind that in practical applications a second
incoherent beam with $\theta^{'}_{\mathrm{c}}=-\theta_{\mathrm{c}}$ is needed in
order to cancel the in-plane contribution of the non-chiral forces
\cite{rukhlenko_completely_2016}. We follow a similar approach when calculating
the comparison force terms for the PW setup, i.e. we employ a single circularly
polarized plane wave at $\lambda=\lambda_{\mathrm{c}}$ in a
$n_{\mathrm{water}}=1.33$ medium, and remember that we need a second incoherent
contribution for any practical application \cite{rukhlenko_completely_2016}.
Finally, for the comparison with the EW force term, we illuminate a glass
$n_{\mathrm{inc}}=1.53$ prism at $\lambda=\lambda_{\mathrm{c}}$ and
$\theta=\theta_{\mathrm{c}}$ with a TM polarized plane wave and monitor the
induced chiral force $30$~nm above the prism water interface
\cite{hayat_lateral_2015}.
\begin{figure}[t!]
\begin{center}
\includegraphics[]{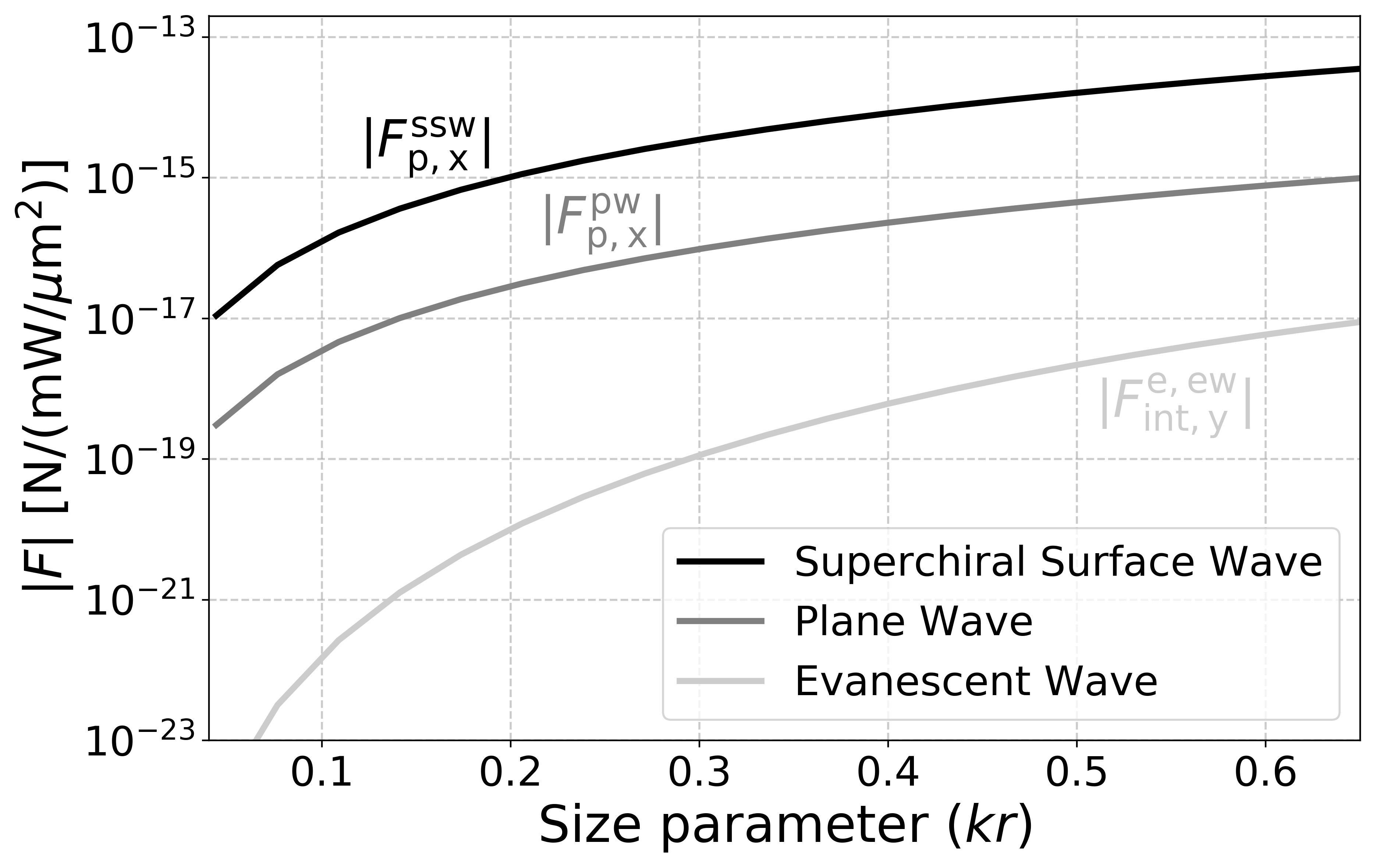}
\caption{\label{fig:force_vs_size} Chiral optical forces, normalized to the
incident power, vs size parameter  for the superchiral surface wave, plane wave
and evanescent wave configurations.}
\end{center}
\end{figure}
Figure \ref{fig:force_vs_size} compares the $|F\mathrm{^{ssw}_{p,x}}|$ and
$|F\mathrm{^{pw}_{p,x}}|$ radiation pressure force terms for SSW and PW setups
against the EW $|F\mathrm{^{e,ew}_{int,y}}|$ force term, as depicted in Figure
\ref{fig:scheme}. The enantioselective force generated by the SSW configuration
is between one and two orders of magnitude larger than that obtained with the
plane wave configuration with the same incident power. The gap becomes larger if
the comparison is made with the transverse interaction terms, reaching a ratio
as large as 5 orders of magnitude towards smaller particle sizes. It is
nevertheless interesting to note that the EW force term display a stronger size
dependency through the $\mathrm{Re}\{ \chi \alpha^{*}_{\mathrm{e}} \}$ factor.
It thus becomes a viable alternative towards the larger side of the size
spectrum, taking also in account the fact that the EW approach requires a single
incident beam setup \cite{hayat_lateral_2015}. Overall, the forces per unit of
incident power achievable with the SSW approach range from
$|F\mathrm{^{ssw}_{p,x}}| \sim 10^{-17}$~N/(1~mW/1$\mu$m$^{2}$) for $r=2$~nm to
$|F\mathrm{^{ssw}_{p,x}}| \sim 10^{-13}$~N/(1~mW/1$\mu$m$^{2}$) for $r=30$~nm particles, making it a
viable solution for all-optical enantiomer separation.

To further investigate the properties of SSW based enantioselective forces, we
examine the 1DPC platform performance for different illumination conditions. To
do so we choose an $r=5$~nm chiral sphere and, as before, we position it 30 nm
above the 1DPC surface.
\begin{figure}[t!]
\begin{center}
\includegraphics[]{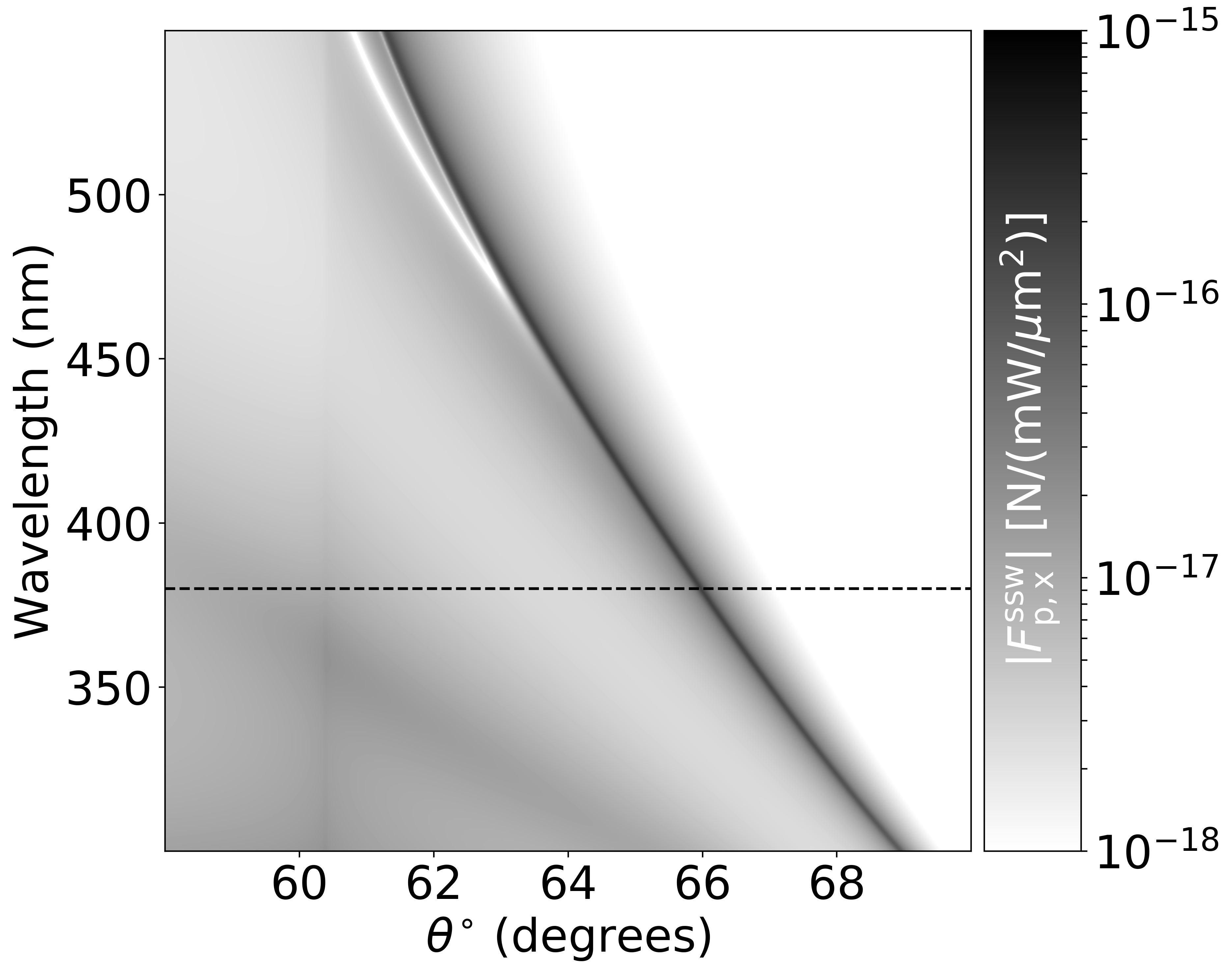}
\caption{\label{fig:force_map} Optical force map, normalized to the incident
power, for the superchiral surface wave configuration, calculated for a $5$~nm
chiral sphere 30~nm above the 1DPC surface. The thin dashed line indicates an
angular slice taken at constant wavelength $\lambda_{\mathrm{c}}=380$~nm.}
\end{center}
\end{figure}
Figure \ref{fig:force_map} reports the map for the $|F\mathrm{^{ssw}_{p,x}}|$
force term for wavelengths in the $\lambda \sim300-550$~nm range and incident
angles in the $\theta \sim58-70^{\circ}$ range. The density plot reveals that
the maximum chiral force is obtained along a diagonal line cutting the map from
the top-left to the bottom-right corner. This maximum force line corresponds to
the superposition of the TE and TM Bloch surface waves dispersion relations or,
in other words, the line follows the dispersion relation of the SSW \cite{pellegrini_chiral_2017}. This
approach therefore provides large chiral forces on a wide energy spectrum,
ranging from the UV to the visible range, at the only expense of tuning the
incident illumination angle to match the SSW excitation conditions. Outside the
dispersion relation region, the obtained forces are on average one to two orders
of magnitude smaller, becoming comparable to those obtained with the simpler
plane wave setup.\

\begin{figure}[t!]
\begin{center}
\includegraphics[]{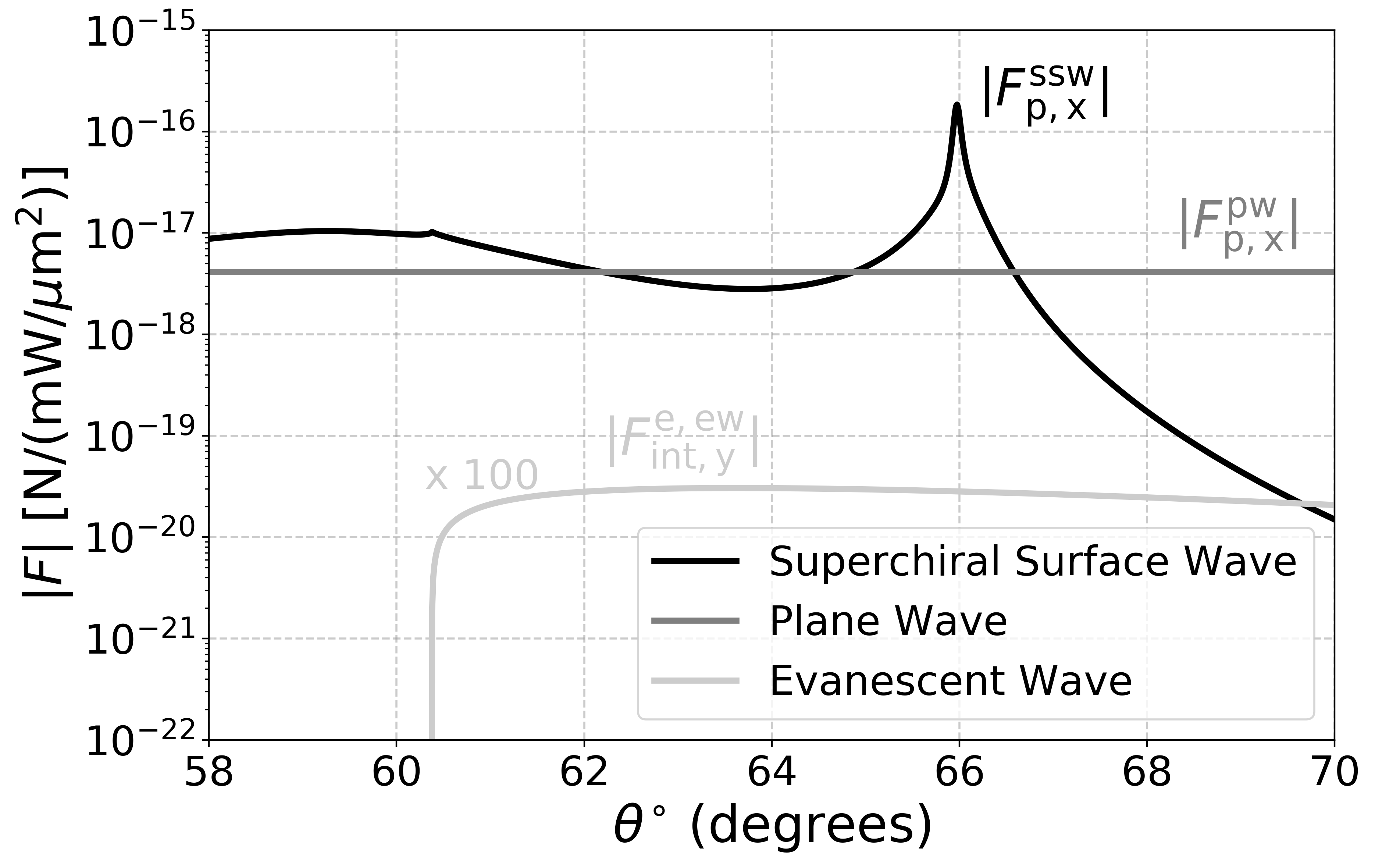}
\caption{\label{fig:force_cut} Chiral optical forces, normalized to the incident
power, vs illumination angle for the superchiral surface wave, plane wave and
evanescent wave configurations. The evanescent wave force term is multiplied by
a factor of $100$, and vanishes below the critical angle.}
\end{center}
\end{figure}

We can obtain further insight in the $|F\mathrm{^{ssw}_{p,x}}|$ term behavior by
slicing the force map at the $\lambda_{\mathrm{c}}=380$~nm wavelength. Figure
\ref{fig:force_cut} reports the obtained force angular spectrum, and compares it
with the force resulting from the alternative PW and EW approaches. Figure
\ref{fig:force_cut} clearly shows that, at the coupling angle around
$\theta_{\mathrm{c}} \sim 66^{\circ}$, SSWs provide chiral force enhancements of
about two orders of magnitude. It is also apparent that, in this size range
($r=5$~nm), evanescent wave forces do not represent a valid alternative for
all-optical enantioseparation, and even more so when dealing with small chiral
molecules. We finally note that the SSW angular force spectrum displays a less
evident feature around $\theta \sim 60.3^{\circ}$, i.e. in correspondence of the
critical angle. The feature indicates that evanescent waves excited with
circularly polarized light can generate radiation pressure chiral forces
($|F\mathrm{^{eva}_{p,x}}|$, along the $x$-axis) that, while being more than one
order of magnitude smaller than the SSW counterpart, are still larger than those
obtained with simple plane waves.\

As a last investigation, we want to study the lateral diffusion of the $r=5$~nm
chiral target, assuming that it is suspended in water, inside a microfluidic
channel, under the influence of the purely chiral forces described above. The
diffusion of the pure, non-interacting enantiomers is described by the
Fokker-Plank equation \cite{rukhlenko_completely_2016,jones_optical_2015},
\begin{equation}
\frac{\partial \rho}{\partial t}=D \frac{\partial^{2} \rho}{\partial x^{2}} - v \frac{\partial \rho}{\partial x}
\label{eq:fokker}
\end{equation}
with vanishing flux boundary conditions at the channel walls
\begin{equation}
x_{0} \frac{\partial \rho}{\partial x}\bigg\rvert_{x=0}=\rho(0,t), \ x_{0} \frac{\partial \rho}{\partial x}\bigg\rvert_{x=L}=\rho(L,t)
\label{eq:boundary}
\end{equation}
and uniform enantiomer distribution $\rho(x,0)=\rho_{0}$ at the time $t=0$. In
the equations, $\rho(x,t)$ is the enantiomer concentration,
$D=\frac{k_{\mathrm{B}}T}{6 \pi \eta r}$ the diffusion coefficient, $v=\frac{F
D}{k_{\mathrm{B}}T}$ the drift velocity under a force $F$, $k_{\mathrm{B}}$ the
Boltzmann's constant, $T$ the solution temperature, $\eta$ the fluid viscosity
and $L$ the microfluidic channel width. It is interesting to note that an
intrinsic spatio-temporal scale emerges from the Fokker-Plank equation, where
the spatial evolution happens on a natural length scale defined as $x_{0} =
D/v$, while for the temporal evolution we can define $t_{0}=D/v^{2}$. Starting
from $t_{0}$ we can finally obtain an effective time constant for the enantiomer
diffusion process with $\tau_{\mathrm{eff}} \sim \tau = 4 t_{0}$ if $L \gg 2 \pi
x_{0}$ and $\tau_{\mathrm{eff}} \sim \tau_{1} = L^{2}/(\pi^2 D)$ if $L \ll 2 \pi
x_{0}$.
\begin{figure}[t!]
\begin{center}
\includegraphics[]{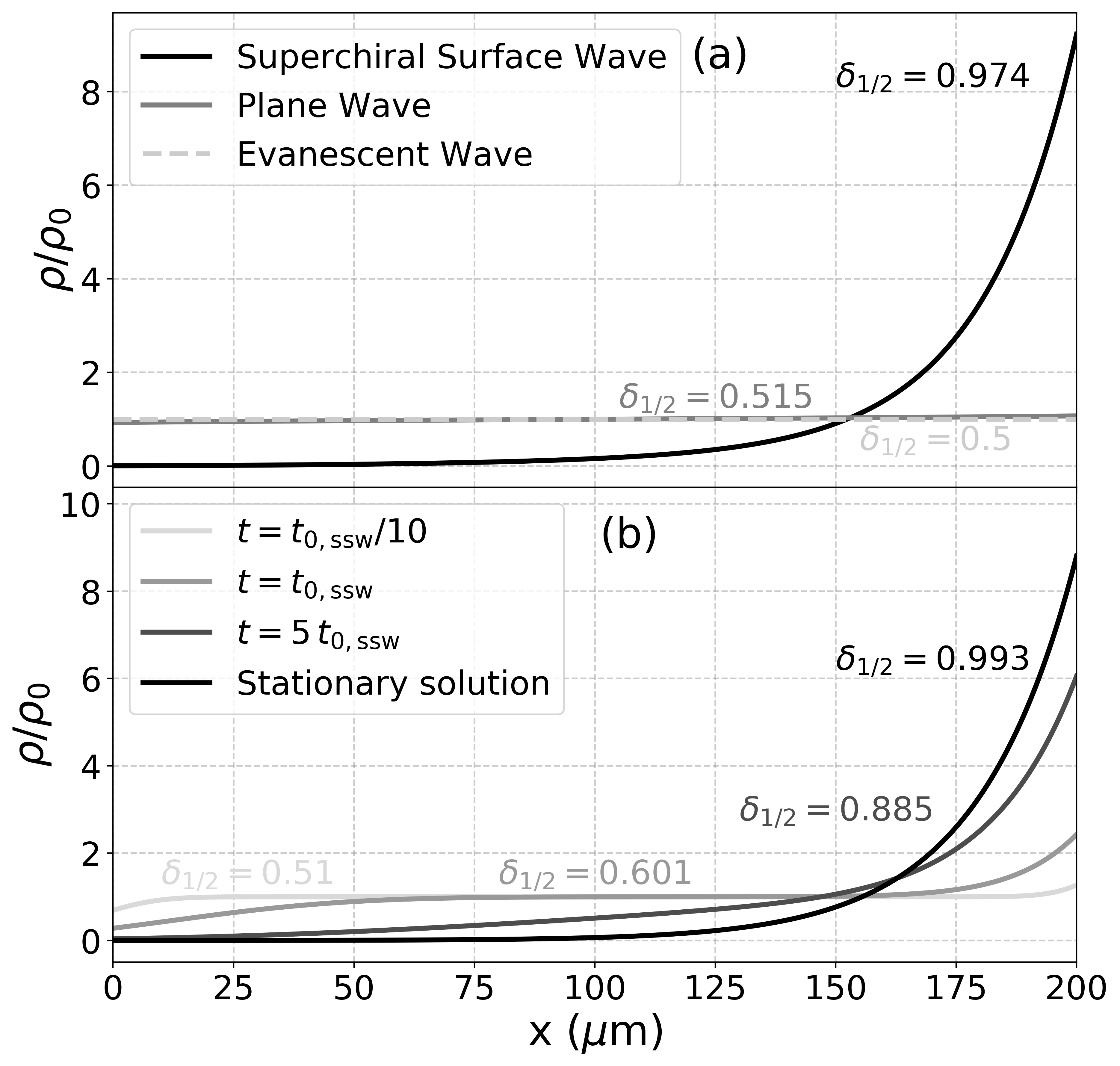}
\caption{\label{fig:diffusion} (a) Enantiomer ($r=5$~nm sphere) concentration
profiles at the time $t=90$~s for the superchiral surface wave, plane wave, and
evanescent wave configurations, where the $\delta_{1/2}$ symbols stands for the
amount of enantiomer contained to the right hand side ($x>100 \, \mu$m) of the
microfluidic channel. (b) Temporal evolution of the enantiomer concentration for
the superchiral surface wave configuration.}
\end{center}
\end{figure}
To calculate the diffusion profiles, we choose a room temperature condition of
$T=293.15$~K, the corresponding water viscosity $\eta = 10^{-3}$~Pl, and an
incident power of 1~mW/1$\mu$m$^{2}$, which is a fairly typical value for
optical tweezers setups \cite{jones_optical_2015}. Consequently, for the
constant force term, we choose the maximum values reported in
Fig.\ref{fig:force_cut}, obtaining $|F\mathrm{^{ssw}_{p,x}}| \sim 2 \cdot
10^{-16}$~N, $|F\mathrm{^{pw}_{p,x}}| \sim 4 \cdot 10^{-18}$~N and
$|F\mathrm{^{e,ew}_{int,y}}| \sim 3 \cdot 10^{-22}$~N, where in any case we
consider the force applied along the $x$-axis to comply with the notation of
Eq.\ref{eq:fokker}. Figure \ref{fig:diffusion} reports the enantiomer
concentration profiles for an $L=200$~$\mu$m microfluidic channel after a time
$t=90$~s, where the adopted channel width and diffusion time are choosen as a
function of the natural spatial and temporal scales emerging when using the
$|F\mathrm{^{ssw}_{p,x}}|$ force term, with $L \sim 10 \, x_{0,\mathrm{ssw}}$
and $t \sim 10 \, t_{0,\mathrm{ssw}}$. The concentration profiles of
Fig.\ref{fig:diffusion}(a) reveal that only the SSW approach can produce a
sizable enantioseparation, with $97.4\%$ of the enantiomer displaced in the
right hand side of the microfluidic channel ($x>100 \, \mu$m). In this case, the
separation process is further facilitated by the presence of strong gradient
forces along the $z$ direction, pushing the chiral compounds towards the 1DPC
surface, where the $|F\mathrm{^{ssw}_{p,x}}|$ term is larger. On the other hand,
the concentration profiles associated to the PW and EW schemes are virtually
unchanged compared to the initial conditions, since the magnitude of the
associated force terms does not grant a substantial profile change on these
spatio-temporal scales. The temporal evolution of the SSW concentration
profiles, displayed in Fig.\ref{fig:diffusion}(b), further reveal that sizable
deviations form the uniform distribution $\rho_{0}$ can be obtained on the time
scale of a few seconds, to finally reach a $99.3\%$ separation for the
stationary solution. It is also clear that, for many practical purposes, the
$t=90$~s concentration profile can be considered virtually identical to the
stationary one.

In conclusion, we have proposed the use of superchiral surface waves for all-optical enantiomeric separation. Our solution provides forces up to two orders of magnitude larger than those obtained with alternative approaches, allowing for the all-optical separation of chiral targets on otherwise unachievable spatial, temporal and size scales. The suggested 1DPC platform can operate in a wide energy range and is inherently compatible with standard microfluidic setups, thus representing a substantial step towards the all-optical separation and manipulation of chiral molecules and nanoparticles.

% \begin{acknowledgement}
%
% The research leading to these results has received funding
% from the Italian Ministry of Education, Universities and
% Research (MIUR) through the PRIN 2015 program (Project
% No. 2015FSHNCB “Plasmon-enhanced vibrational circular
% dichroism”).
%
% \end{acknowledgement}

% \begin{suppinfo}
%
% This will usually read something like: ``Experimental procedures and
% characterization data for all new compounds. The class will
% automatically add a sentence pointing to the information on-line:
%
% \end{suppinfo}

\bibliography{bibliography}

\end{document}